\PassOptionsToPackage{prologue,table,xcdraw}{xcolor}
\documentclass[sigconf,screen]{acmart}

\usepackage{graphicx}
\usepackage{textcomp}
\usepackage[table,xcdraw]{xcolor}
\usepackage{multirow}
\usepackage[inline]{enumitem}
\usepackage{url}
\usepackage[capitalize,noabbrev]{cleveref}
\usepackage[most]{tcolorbox}

\usepackage[T1]{fontenc}
\usepackage{nicefrac}

\usepackage{siunitx}  
\usepackage{flushend}
\usepackage{dirtytalk}
\usepackage{booktabs}
\usepackage{microtype}

\SetEnumitemKey{midsep}{topsep=3pt, partopsep=0pt}
\newif\iftodos
\todostrue

\newcommand{\Dxy}[0]{$D_{xy}$}
\newcommand{\CI}[0]{$95\,\%\,\text{CI}$}

\iftodos
\newcommand{\PP}[1]{\todo[inline]{#1}}
\newcommand{\GK}[1]{\todo[inline,color=green!40]{GK: #1}}
\newcommand{\WW}[1]{\todo[inline, color=green!40,size=\tiny]{WW: #1}}
\newcommand{\KR}[1]{\todo[inline, color=green!40,size=\tiny]{KR: #1}}
\else
    \newcommand{\PP}[1]{}
    \newcommand{\GK}[1]{}
    \newcommand{\WW}[1]{}
    \newcommand{\KR}[1]{}
\fi

\def\BibTeX{{\rm B\kern-.05em{\sc i\kern-.025em b}\kern-.08em
    T\kern-.1667em\lower.7ex\hbox{E}\kern-.125emX}}

\copyrightyear{2026}
\acmYear{2026}
\setcopyright{cc}
\setcctype{by}
\acmConference[MSR '26]{23rd International Conference on Mining Software Repositories}{April 13--14, 2026}{Rio de Janeiro, Brazil}
\acmBooktitle{23rd International Conference on Mining Software Repositories (MSR '26), April 13--14, 2026, Rio de Janeiro, Brazil}
\acmPrice{}
\acmDOI{10.1145/3793302.3793364}
\acmISBN{979-8-4007-2474-9/2026/04}

\begin{document}

\title{Linux Kernel Recency Matters, CVE Severity Doesn't, and History Fades}

\author[1]{Piotr Przymus}
\orcid{0000-0001-9548-2388}
\affiliation{
   \institution{Nicolaus Copernicus University}
   \city{Toru\'n}
   \country{Poland}}
\email{piotr.przymus@mat.umk.pl}

\author[2]{Witold Weiner}
\orcid{0009-0001-0849-4303}
\affiliation{
   \institution{Nicolaus Copernicus University}
   \city{Toru\'n}
   \country{Poland}}
\email{wweiner@mat.umk.pl}

\author[2]{Krzysztof Rykaczewski}
\orcid{0000-0002-4772-6330}
\affiliation{
   \institution{Nicolaus Copernicus University}
   \city{Toru\'n}
   \country{Poland}}
\email{mozgun@mat.umk.pl}

\author[4]{Gunnar Kudrjavets}
\orcid{0000-0003-3730-4692}
\affiliation{
   \institution{Amazon Web Services$^{\ast}$}
   \city{Seattle}
   \state{WA}
   \postcode{98109}
   \country{USA}}
\email{gunnarku@amazon.com}
\thanks{$^\ast$ Conducting research is not related to Gunnar Kudrjavets' role at Amazon Web Services, Inc. All opinions and statements communicated in this paper are the author's own.}

\begin{abstract}
In 2024, the Linux kernel became its own Common Vulnerabilities and Exposures (CVE) Numbering Authority (CNA), formalizing how kernel vulnerabilities are identified and tracked.
We analyze the anatomy and dynamics of kernel CVEs using metadata, associated commits, and patch latency to understand what drives patching.
Results show that severity and Common Vulnerability Scoring System (CVSS) metrics have a negligible association with patch latency, whereas kernel recency is a reasonable predictor in survival models.
Kernel developers fix newer kernels sooner, while older ones retain unresolved CVEs.
Commits introducing vulnerabilities are typically broader and more complex than their fixes, though often only approximate reconstructions of development history.
The Linux kernel remains a unique open-source project---its CVE process is no exception.
\end{abstract}

%
%
\keywords{Linux kernel, CVE, CNA, CVSS, survival analysis}

\maketitle

\section{Introduction}
\label{sec:intro}
The Linux kernel is one of the most widely deployed open-source projects.
As of October 2025, the Linux kernel comprises over 40 million lines of code~\cite{scc}.
The kernel's security is critical to the global communications and computing infrastructure.
The Linux kernel runs in smart refrigerators and powers space exploration for \textsc{NASA} and {SpaceX}.
Serious security events, such as the widespread exploitation of a zero-day vulnerability in the Linux kernel, have the potential to cause significant negative societal impact.
The consequences of an unpatched security vulnerability in the kernel can rival the scope of the {C}rowd{S}trike incident from 2024~\cite{crowdstrike}.

The industry standard for tracking and communicating vulnerabilities is the CVE program, established in 1999 and maintained by the \textsc{MITRE} {C}orporation~\cite{cve-process}.
The accepted practice in the software industry is to mitigate attacks and maintain the security posture of various systems by continuously patching them against known vulnerabilities.
Based on our anecdotal observations from academia and the industry, CVEs are often stigmatized and have a negative connotation associated with them~\cite{the-stigma-of-cve}.
The common emotional reaction to the CVEs often tends to be \say{we must fix all the CVEs immediately} versus treating them as \say{just a tracking number}~\cite{the-stigma-of-cve}.

On February 13, 2024, the Linux kernel project became its own CNA, and taking direct responsibility for assigning and managing CVEs~\cite{kernel-became-cna}.
As a result of this change, the Linux kernel project started issuing an order of magnitude more CVEs than various entities have reported in the past~\cite{tux_care}.
The transition marked a cultural, as well as procedural, shift from the Linux community's skepticism toward CVEs to an institutionalized vulnerability management process~\cite{gkh-kernel-recipes-2019-talk}.
This milestone enables a systematic study of how vulnerabilities emerge and evolve in the Linux kernel, as the definition of kernel CVEs has expanded beyond security.
We observe that the Linux community's daily active involvement in the kernel's CVE process has been consistent and more involved than in other comparable open-source operating systems (e.g., {FreeBSD}, {NetBSD}).

The increase in the number of kernel CVEs has organizational and procedural consequences for all the entities that maintain or publish custom kernels.
Each kernel provider needs to decide how to modify their existing engineering processes related to triaging incoming CVEs, (potentially) backporting patches, and communicating the impact and risk of CVEs to the kernels' consumers.
All kernel providers will face the same challenges of determining in what order they should expect the Linux kernel community to backport fixes to older {L}ong-{T}erm {S}upport (LTS) kernels, whether CVSS scores and National Vulnerability Database (NVD) ratings matter, and if there is something unique about the commits that cause CVEs to be issued.

This paper analyzes \num{6464} unique CVEs issued since the Linux kernel became CNA to understand how vulnerabilities are introduced, classified, and fixed.
We focus on three aspects:
\begin{enumerate*}[label=(\alph*),before=\unskip{ }, itemjoin={{, }}, itemjoin*={{, and }}]
    \item the structural characteristics of vulnerability-inducing versus fixing commits
    \item the role of CVE metadata (CVSS scores and vectors) in explaining the fix latency
    \item the effect of kernel version age on the speed of patch propagation.
\end{enumerate*}

\subsection{Research Questions}
\label{sec:RQs}
To help the software industry make optimal decisions about their engineering processes related to kernel consumption and maintenance, we structure our investigation around the following research questions:

\begin{itemize}
    \item \textbf{RQ1:} What is the anatomy of vulnerability-inducing versus fixing commits, and what does it reveal about the cost of fixing vulnerabilities?
    \item \textbf{RQ2:} What role does CVSS scores play in the speed of vulnerability fixing in the Linux kernel?
    \item \textbf{RQ3:} How does the kernel version age influence the speed of vulnerability fixes?
\end{itemize}

\paragraph{Definition of cost.}
Throughout the paper, we use the term \emph{cost} solely to denote the \emph{observed magnitude of code changes}, operationalized as lines-of-code (LOC) churn in commits.
In \textbf{RQ1}, we therefore compare vulnerability-introducing and vulnerability-fixing commits only using LOC-based measures (e.g., added/removed lines, files touched) and our line-level change annotations (e.g., \texttt{code}, \texttt{documentation}, \texttt{configuration}, \texttt{tests}).
We do \emph{not} attempt to estimate developer effort, time-to-produce patches, or financial cost.

Together, these questions help reveal whether vulnerability management in the kernel follows measurable, predictable patterns or is dominated by subsystem-specific dynamics.

\subsection{Contributions and Replication}
\label{sec:contribs}
Our contributions are threefold:
\begin{enumerate}
  \item We provide an empirical analysis of thousands of kernel CVEs, introducing and fixing commits and CVSS metadata.
  \item We apply survival analysis to model vulnerability fixing latency, identifying factors that correlate with longer-lived CVEs.
  \item We release a replication package~\cite{Przymus2026cve} including data extraction scripts and curated CVE-commit mappings.
\end{enumerate}

\section{Background and Related Work}
\label{sec:related}
\subsection{The {Linux} Kernel Development Model}

The Linux kernel development model is a well-documented process that has followed a similar pattern for more than three decades~\cite{linux-kernel-process}.
The project adheres to the core principles of open-source software development.
The kernel development model uses distributed development, open communication, and patch-based changes that kernel contributors discuss via mailing lists~\cite{jiang_2012,tan_2019,israeli_2010}.
Although the Linux kernel is an open-source project, the decision-making process within the Linux kernel community is hierarchical.
Linus Torvalds acts as the primary coordinator for the Linux kernel and the final arbiter of all decisions.
He relies on a core group of trusted lieutenants (e.g., Greg Kroah-Hartman, Ingo Moln\'{a}r, Andrew Morton) who maintain various branches, releases, and subsystems.
The code contributions to the kernel are evaluated based on their merit.

The unit of architectural granularity in kernel development is a subsystem.
A subsystem is a specific, logical part of the kernel (e.g., memory management, scheduling, or specific hardware drivers).
One or more maintainers oversee each subsystem.
Maintainers are senior kernel engineers who are responsible for the overall quality of a specific subsystem.
Developers submit each contribution to the kernel as a patch.
A patch (also known as a code review request, diff, or pull request) is the most granular unit of change in the kernel code.
Maintainers review each incoming patch and decide whether to accept or reject it.
The maintainers act as gatekeepers for code quality and the code meeting kernel standards.

As with other open-source projects, developers choose what tasks they work on.
Developers choose the tasks to work on based on their interests and expertise.
In the context of the kernel, those tasks can include fixing defects, developing new features, or implementing drivers to support new hardware.
The only exception here is kernel developers who work for commercial entities.
In those situations, the business and project needs dictate what the engineers work on.
The system operates on a high degree of developer autonomy and meritocracy.

In most cases, developers handle vulnerabilities the same way they handle other defects.
Historically, some embargoed vulnerabilities have been an exception to this process.
There is very little bureaucracy and coordination needed to develop a fix for a security vulnerability.
If a kernel security vulnerability is confirmed, a patch is typically developed, reviewed, and merged at an accelerated pace (hours, days).
The patch itself follows the same process, in which a regular subsystem maintainer responsible for the code must accept the code changes.

\subsection{CVEs in the Kernel Ecosystem}

\begin{flushright}
    \emph{\say{And people shouldn't "fear" a CVE, it's just a hint of "hey, here's a bugfix that might pertain to you!" type of a signal. It's up to the receiver to determine if it does or not.}}
\end{flushright}
\begin{flushright}
--- Greg Kroah-Hartman~\cite{cve-is-just-a-signal}.
\end{flushright}

Historically, CVEs have referred to security flaws, such as exploitable buffer overflows, leaks of private data, and zero-day vulnerabilities.
The Linux kernel community holds the viewpoint that security defects are a subset of defects that do not need distinctive attribution---\say{bug is a bug is a bug}~\cite{gkh-kernel-recipes-2019-talk,gkh-kernel-recipes-2023-talk}.
In 2019, {Greg} {Kroah-Hartman} stated that \say{[F]or the Linux kernel, CVEs do not work at all given the rate of fixes being applied and rapidly backported and pushed to users through a huge variety of different ways}~\cite{gkh-kernel-recipes-2019-talk}.
That viewpoint changed on February 13, 2024, when the Linux kernel project became a CNA to ensure that the kernel team has control over how the CVEs are handled in the kernel~\cite{kernel-became-cna}.
This organizational change triggered a passionate debate in the Linux kernel community~\cite{lwn-cve-article}.

As of 2025, CVE assignment in the kernel is public and straightforward.
The Linux kernel CVE team reviews \emph{each} commit that goes into the stable kernel release~\cite{gkh-talk-2024-suddenly}.
A team of 3--4 senior kernel engineers determines if a commit meets the established criterion for the Linux kernel CVE and votes on the commit's classification~\cite{gkh-talk-cves-are-alive}.
The guiding principle is that \say{\dots anything that fixes a weakness
or an unexpected result should be assigned a CVE}~\cite{official-kernel-cve-definition}.

The Linux kernel CVE team does not intentionally assign any severity to the CVEs.
The primary reason is that the CVE severity depends on a specific use case.
A wide range of contexts and scenarios utilize the kernel---everything from embedded systems to warehouse-scale computing.
However, the Linux kernel CVE team has no insight into all the possible usage scenarios.

Based on anecdotal observations, the industry primarily uses CVSS scores from the NVD as a universal indicator of a vulnerability's \emph{risk} or \emph{technical severity}.
The ambiguity in the interpretation comes from differing guidance.
For example, in the United States, the vulnerability scanning guidelines from the {Federal} {Risk} and {Authorization} {Management} {Program} specify that \say{\dots\ the CVSSv3 base score must be used as the original risk rating}~\cite{fedramp-scanning}.
However, the CVSS 4.0 standard focuses on the severity.
It states that consumers need to combine different CVSS metric values \say{\dots\ specific to their use of the vulnerable system to produce a score that provides a more comprehensive input to risk assessment specific to their organization}~\cite{cvss-4.0-standard}.
Additionally, various kernel distributions (e.g., Fedora, Red Hat, Ubuntu) provide their own assessments of each CVE.
Those assessments are specific to the configuration and intended use of a specific distribution.

\subsection{Empirical Studies of Vulnerabilities in OSS}

Survival analysis has been applied to the lifecycles of vulnerabilities and defects by numerous researchers in empirical studies.

Wedel et al.~\cite{DBLP:conf/esem/WedelJG08} employed survival analysis to analyze the Eclipse project.
They found that code size was the most significant factor for defect duration in a release.

In a survival analysis of defects across four projects~\cite{DBLP:conf/wcre/CanforaCCP11},
the authors linked the lifespans of defects to specific code syntax and identified those related to long-lived defects.

A survival analysis of issues was carried out in a study of selected GitHub projects~\cite{DBLP:conf/socinfo/JarczykGJBW14}.
Its authors analyzed \num{2000} projects for issue survival,
using binomial regression against approximately a dozen basic project property features.

Survival of vulnerabilities in Android was the subject of the study~\cite{DBLP:conf/msr/VasquezBE17}.
This work indicates that more severe vulnerabilities are those that live longer.
The authors used a schema of vulnerability life that includes an attempt to identify when the vulnerability was \emph{introduced},
rather than only the \emph{reporting} time.

Householder analyzed public exploit availability
in ExploitDB and the Metasploit framework between 2013 and 2020 on \num{75807}
software vulnerabilities with assigned CVE IDs~\cite{DBLP:conf/uss/HouseholderCNWS20}.
The authors used survival analysis via the Kaplan-Meier estimator,
with exploit publish date as the endpoint, to discover how CVE ID features influence exploit preparation~\cite{kaplan-meier}.

Nappa et al.~\cite{DBLP:conf/sp/NappaJBCD15} researched the deployment of software patches to hosts.
They used survival analysis on combined patch delay and patch deployment, defined as the time between vulnerability disclosure and the time of patch arrival to user hosts (software update).
The authors analyzed 10 popular client-side applications affected by \num{1593} vulnerabilities,
using data from 8.4 million hosts over a span of five years.

Przymus et al.~\cite{Przymus_2023} present a large-scale analysis of CVE lifetimes using survival analysis across many open-source projects, examining language, project, and CVE-intrinsic risk factors via survival modeling.

Frei et al.~\cite{DBLP:conf/sigcomm/FreiMFP06} conducted an analysis of 14,000 vulnerabilities published between 1996 and 2006
to determine the dates of discovery, disclosure, exploit, and patch.
They correlated CVE entries with \num{80000} additional sources,
and found that 95\,\% of exploits are available within a month of disclosure,
while between 55\,\% and 85\,\% of patches are available on the same date or within 30 days.

Kudrjavets~\cite{Kudrjavets2025PatchMeIfYouCan} discussed the need for systematic, data-driven analyses of CVE management within the Linux kernel after it became a CNA.
Their position paper outlined challenges related to patch issuance velocity, branch-level propagation, and the growing number of CVEs affecting maintainers.
Our study directly addresses these challenges by providing the first survival-based analysis of kernel CVEs, linking vulnerability metadata, patch characteristics, and propagation behavior across versions.

While prior studies examined projects such as Android, Eclipse, or client-side applications, none focused on the Linux kernel after it became a CNA.
Our study therefore, extends the existing literature by connecting vulnerability lifecycle modeling with kernel-specific development practices.

\section{Methodology}
\label{sec:method}

\textbf{Survival analysis.}
Survival analysis models time-to-event data, where the event of interest in our case is a CVE being fixed~\cite{liu2012survival}.
The survival function $S(t)$ is the probability that a CVE remains unfixed at time $t$, with $S(t)=1-F(t)$ if $F$ denotes the cumulative distribution function of the event time.

We estimate the survival function via the Kaplan--Meier estimator with right-censoring:
\begin{equation}
    \hat{S}(t)=\prod_{t_k\le t}\left(1-\frac{d_k}{n_k}\right),
\end{equation}
where \(d_k=\sum_i \mathbb{I}[T_i=t_k,\,\Delta_i=1]\) and \(n_k=\sum_i \mathbb{I}[T_i \ge t_k]\).

For each observation $i$ at the (LTS branch, CVE) level, we define the start time $T_{s,i}$ as the timestamp of the vulnerability-introducing commit identified by \texttt{dyad}, and the end time $T_{e,i}$ as the timestamp of the \emph{earliest} fixing (backported) commit observed in that LTS branch.
We measure time-to-fix in days as $Y_i = T_{e,i} - T_{s,i}$ (computed from commit timestamps and expressed in days).
To accommodate incomplete fix histories, we use the event indicator $\Delta_i \in \{0,1\}$ and observed time $T_i = \min(T_{e,i}, C)$, where $C$ is the end of the study window; observations with no fix by $C$ are treated as right-censored ($\Delta_i=0$).

Because of how the Linux kernel community tracks CVEs, most vulnerabilities can be linked to both introducing and fixing commits.
While this enables time-based analysis, the origin times $T_{s,i}$ should be viewed as approximations rather than exact introduction dates (see~\Cref{subsec:spot-check}).
Introducing commits may be missing for pre-\textsc{Git} development history, and inferred links can be affected by commit squashing, refactoring, or delayed CVE assignment.
Therefore, absolute lifetimes should be interpreted with this limitation in mind.
Conversely, we exclude cases where either the introducing or fixing commit timestamp is missing to ensure consistent time-to-fix estimation.

Dates from CVE metadata alone are unreliable because assignments may be retroactive and publication dates can lag behind the actual fixes.
Therefore, we rely primarily on data from the Linux kernel's CVE repository~\cite{vulns-git-repo}.
Because fixes propagate at different speeds across branches, the same CVE may exhibit distinct lifetimes in different LTS versions.

\textbf{Association and confidence.}
We assess the association between explanatory factors (e.g., kernel branch)
and CVE time-to-fix using Somers' rank association statistic $D_{xy}$,
which ranges from $-1$ (perfect discordance) to $+1$ (perfect concordance)~\cite{newson2010comparing}.
Because the data are right-censored, $D_{xy}$ is computed on comparable pairs only,
i.e., pairs for which the ordering of event times is observable.
To quantify uncertainty, we compute $95\,\%$ confidence intervals via
bootstrap resampling at the CVE level (sampling CVEs with replacement),
recalculating the survival estimates and $D_{xy}$ in each replicate.

\subsection{Commit Anatomy Analysis}
\label{sec:anatomy}

To analyze structural differences between vulnerability-introducing and vulnerability-fixing commits,
we use \emph{PatchScope}, a framework for automatic line-level annotation of code diffs~\cite{patchscope}.
PatchScope combines lexical analysis, file metadata, and path-based heuristics to classify each modified line
into categories such as \texttt{code}, \texttt{configuration}, \texttt{documentation}, and \texttt{tests}.
This fine-grained approach goes beyond traditional coarse metrics
(e.g., number of files or lines changed) and enables us to quantify whether fixes primarily involve
code edits, configuration updates, documentation adjustments, or test improvements.

We apply the same annotation procedure to both vulnerability-introducing and vulnerability-fixing commits,
allowing a direct comparison of their scope and structure.
All analyses use PatchScope with default settings.
For each annotated patch, we compute summary statistics, including the number of modified files, the total number of added/removed lines, and the added/removed lines per annotation category.

Some of the counted additions and deletions correspond to line edits where a line is both removed and re-added in modified form.
While we do not distinguish such fine-grained edits separately,
this simplification does not materially affect our analysis.
The goal is to capture the overall scale and structure of changes rather than their syntactic microdynamics.

\section{Data Collection}
\label{sec:data}

\begin{figure}[htpb]
    \centering
    \includegraphics[width=0.4\textwidth]{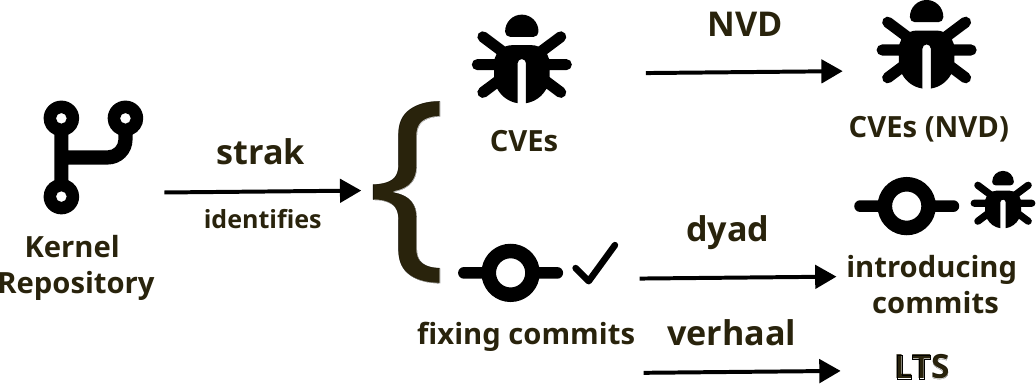}
    \caption{Data collection pipeline in which CVEs and fixing commits are extracted from the kernel repository, vulnerability-introducing commits are identified via line-level blame, commits are mapped to LTS releases, and NVD data is used only for metadata enrichment.}
    \label{fig:tool}
\end{figure}

We integrate multiple data sources to construct a unified dataset combining CVE records issued since the Linux kernel became a CNA (including CVSS\,v3.1 vectors), the complete kernel commit history from both mainline and LTS branches, and explicit links between CVEs and commits extracted from references in CVE entries and kernel changelogs.

\Cref{fig:tool} presents the overall process and tools used. CVE information was collected through a four-phase extraction pipeline using specialized Linux kernel tools:
\texttt{strak}~\cite{vulns-git-repo} (CVE-fix analyzer), \texttt{dyad}~\cite{vulns-git-repo} (vulnerability-fix pair finder), \texttt{verhaal}~\cite{GregkhVerhaalLinux} (commit-to-release mapper), and the NVD API.
The analysis covered six LTS branches (6.12, 6.6, 6.1, 5.15, 5.10, 5.4), spanning approximately \num{1000} kernel versions.
\texttt{strak} identified CVE-fixing commits via \texttt{git log} pattern matching, \texttt{dyad} inferred vulnerability-introducing commits for timeline reconstruction, \texttt{verhaal} mapped commits to release cycles, and the NVD API supplied metadata enrichment.

\begin{figure*}[htb]
\centering
\includegraphics[width=\linewidth]{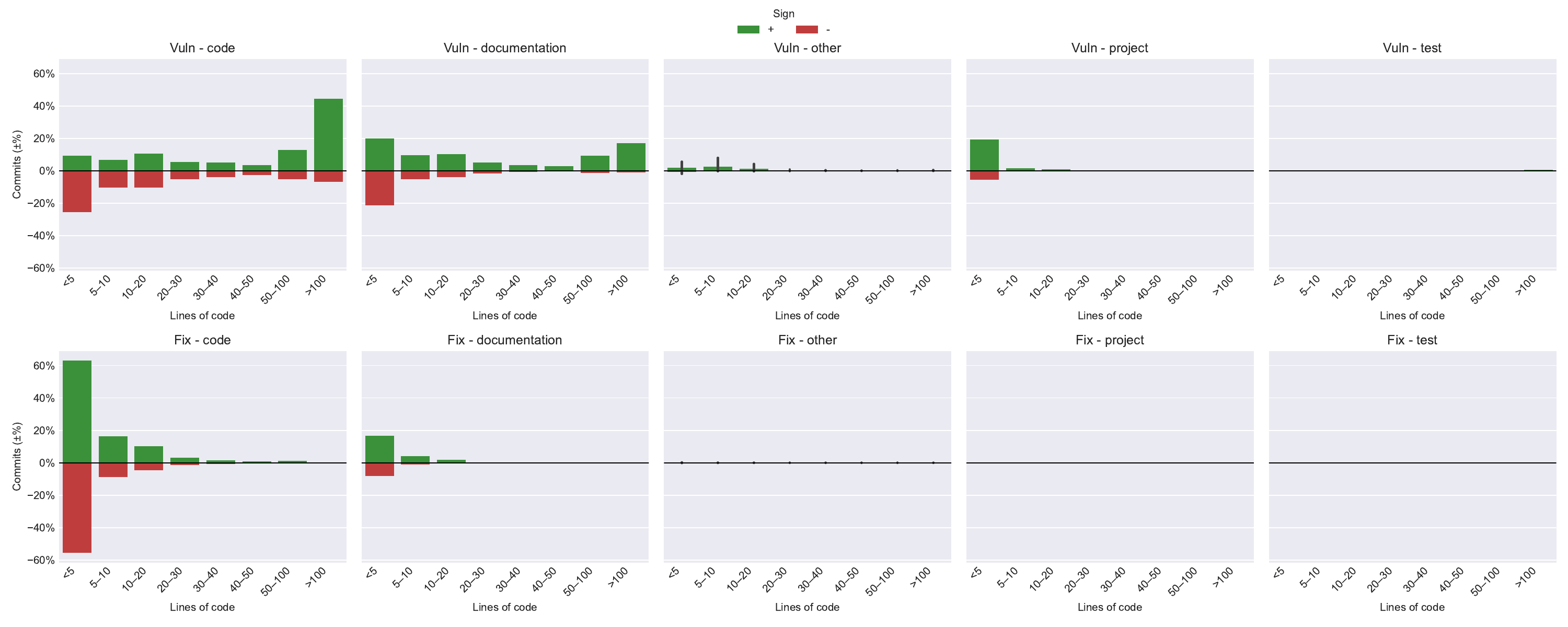}
\caption{
Distribution of commit sizes by annotation type for vulnerability-introducing (top row) and fixing (bottom row) commits.
Each panel corresponds to one annotation type (e.g., code, documentation, tests), and bars are split by change sign: additions (\texttt{+}, green) and deletions (\texttt{-}, red).
The x-axis bins commits by the number of changed lines in that type/sign (as indicated by the bin labels).
The y-axis reports the \emph{percentage of commits} in the corresponding class (Vuln/Fix) that fall into each bin (i.e., counts normalized by the total number of commits in the row).
Commits with zero or missing changes for a given type/sign are not assigned to any bin and therefore do not contribute to the bars (empty changes are not shown).
Note that a single commit may touch multiple annotation types, hence it can contribute to multiple panels; consequently, percentages do not sum to 100\% across panels.
}
\label{fig:commit-anatomy}
\end{figure*}

\Cref{tab:lts-coverage-joined} provides an overview of our dataset coverage across six actively maintained Linux LTS kernel branches.
In total, we identified \num{6464} unique CVEs, of which CVSS\,v3.1 metadata was available for 58\,\%.

We paired each CVE with the corresponding Linux kernel versions in which it was observed, yielding about \num{45900} (CVE, Linux version) pairs across all LTS branches.
Because individual vulnerabilities often affect multiple versions, this pairing captures their propagation across releases.
In most cases (73\,\%), both vulnerability-introducing and fixing commit timestamps were identified, enabling precise measurement of time-to-fix intervals and resulting in \num{32857} LTS-CVE instances.
Temporal and CVSS metadata co-occurred in \num{19693} cases (43\,\%), providing a joint basis for survival and severity analysis.
All subsequent analyses operate at the LTS-CVE level, treating each (LTS, CVE) pair as a distinct observation to capture branch-specific patching behavior, including differences in fix latency and backport schedules for the same vulnerability.

\begin{table}[htb]
  \centering
\caption{Dataset coverage across Linux kernel LTS branches, including CVE counts and metadata availability. \emph{Notes:} ``Dates'' indicates both the vulnerability-introducing and fixing commit timestamps are known. ``Meta'' refers to available CVSS\,v3.1 metadata. ``D$\cap$M'' indicates cases where both Dates and Meta are available.}
  \label{tab:lts-coverage-joined}
  \begin{tabular}{rrrrr}
    \toprule
       & \multicolumn{2}{c}{\rotatebox{0}{Unique}} & \multicolumn{2}{c}{\rotatebox{0}{Releases}} \\
      \cline{2-5}
     LTS & \rotatebox{0}{CVE} & \rotatebox{0}{Meta}& \rotatebox{0}{Dates}& \rotatebox{0}{D$\cap$M}\\
    5.4  & \num{2402} & \num{1488} & \num{5159} & \num{3327} \\
    5.10  & \num{3381} & \num{2131} & \num{6933} & \num{4399} \\
    5.15  & \num{3881} & \num{2291} & \num{7192} & \num{4376} \\
    6.1  & \num{3249} & \num{1951} & \num{5820} & \num{3524} \\
    6.6  & \num{3666} & \num{2219} & \num{5267} & \num{3269} \\
    6.12  & \num{1953} & \num{605} & \num{2486} & \num{798} \\
    \midrule
    \textbf{Total}  & \textbf{\num{6464}} & \textbf{\num{3788}} & \textbf{\num{32857}} & \textbf{\num{19693}} \\
    \bottomrule
  \end{tabular}
\end{table}

\begin{table}[htb]
  \centering
\caption{Time-to-fix distribution of CVEs in Linux Kernel. Buckets represent time-to-fix intervals in days.}
  \label{tab:lts-coverage-times}
  \begin{tabular}{lrrrrrr}
    \toprule
    & \multicolumn{5}{c}{\textbf{Time-to-fix buckets (days)}} & \\
    \cmidrule(lr){2-6}
    LTS   &
    \multicolumn{1}{c}{\textless{}7} &
    \multicolumn{1}{c}{7--30} &
    \multicolumn{1}{c}{31--90} &
    \multicolumn{1}{c}{91--365} &
    \multicolumn{1}{c}{\textgreater{}365} &
    \multicolumn{1}{c}{\textbf{Total}} \\
    \midrule
    5.4  & \num{6} & \num{57} & \num{131} & \num{183} & \num{4782} & \num{5159} \\
    5.10  & \num{26} & \num{63} & \num{156} & \num{348} & \num{6340} & \num{6933} \\
    5.15  & \num{25} & \num{83} & \num{179} & \num{417} & \num{6488} & \num{7192} \\
    6.1  & \num{19} & \num{73} & \num{141} & \num{331} & \num{5256} & \num{5820} \\
    6.6  & \num{14} & \num{69} & \num{129} & \num{504} & \num{4551} & \num{5267} \\
    6.12  & \num{5} & \num{35} & \num{79} & \num{427} & \num{1940} & \num{2486} \\
    \bottomrule
  \end{tabular}
\end{table}

The distribution of fix times (\Cref{tab:lts-coverage-times}) is heavily skewed.
Over 89\,\% of vulnerabilities are fixed more than one year after their introduction.
Only a small fraction (1.4\,\%) are resolved within a month, and same-week fixes are extremely rare (0.14\,\%).
The statistics are reported per LTS-CVE instance, not per unique CVE, meaning that the same CVE may appear multiple times, once for each branch where it was observed.
This approach reflects the true operational burden on maintainers, as each branch requires its own patch integration and validation.

Each commit introducing and fixing commit was further preprocessed by automatically classifying modified lines into four categories (\texttt{code}, \texttt{documentation}, \texttt{configuration}, and \texttt{tests}), enabling structural comparisons between vulnerability-introducing and fixing commits (see ~\Cref{sec:anatomy}).

Having established a consolidated dataset linking CVEs, kernel versions, and commit-level timelines,
we now turn to the analysis.

\section{Results}
\label{sec:results}

Different aspects of this study require viewing the data at different levels of granularity.
When analyzing \textbf{CVSS\,v3.1 characteristics} (\cref{sub:CVSS vectors analysis}), we focus on \emph{unique CVEs} and their associated metadata, reflecting how vulnerabilities are scored in external databases such as NVD.
For \textbf{survival and fix-latency analyses} (\Cref{sub:cve_score}, \Cref{sub:Fixes across LTS versions}), we operate at the \emph{LTS-CVE level}, where each \emph{(CVE, Linux version)} pair represents a distinct observation.
This perspective captures branch-specific patching dynamics, including differences in backport schedules and fix latencies for the same vulnerability across releases.
Finally, for the \textbf{analysis of introducing and fixing commits} (\Cref{sub:commit-anatomy}, \Cref{subsec:spot-check}), we examine \emph{commit pairs} linked to each \emph{(CVE, Linux version)} pair.
If a commit has already been analyzed in another branch, it is deduplicated to avoid repeated counting while preserving its association with multiple fixes.

\subsection{Anatomy of Commits: Change Size and Type}
\label{sub:commit-anatomy}

To understand how vulnerabilities are introduced and resolved in the Linux kernel, we analyzed the structure of commits in terms of the number and type of lines changed.
Our goal is to capture the asymmetry between vulnerability-introducing commits and their corresponding fixes, both in size and in affected subsystems.

\Cref{tab:commit-statistics} summarizes the per-commit change characteristics across all observed commits.
Vulnerability-introducing commits are substantially larger: they typically affect multiple files ($M = 3$), introduce significantly more lines ($M = 94$ of added lines), and involve broader modifications.
In contrast, fixing commits are much smaller and more focused: 75\,\% modify no more than 9 lines, and the vast majority touch only a single file. Outliers are discussed in~\cref{subsec:spot-check}.

\begin{table}[h]
\centering
\caption{Summary statistics for vulnerability-introducing and fixing commits. Each column reports per-commit counts of changed files or lines (\texttt{+} added, \texttt{-} removed).}
\label{tab:commit-statistics}
\begin{tabular}{lrrrrrr}
\toprule
 & \multicolumn{3}{c}{Introducing}& \multicolumn{3}{c}{Fixing} \\
\cmidrule(lr){2-4}
\cmidrule(lr){5-7}
 & Files & + Lines & - Lines & Files & + Lines & - Lines \\
\midrule
Mean & 4 & 575 & 42 & 1 & 9 & 6 \\
Std & 19 & \num{4650} & 170 & 1 & 17 & 14 \\
Min & 1 & 1 & 1 & 1 & 1 & 1 \\
25\,\% & 1 & 14 & 3 & 1 & 2 & 1 \\
50\,\% & 2 & 56 & 9 & 1 & 4 & 2 \\
75\,\% & 4 & 239 & 27 & 1 & 9 & 6 \\
Max & {920} & \num{204645} & \num{3430} & {27} & 423 & 664 \\
\bottomrule
\end{tabular}
\end{table}

\Cref{fig:commit-anatomy} provides further insight by showing the distribution of commits bucketed by the number of lines added or removed, and grouped by annotation type (e.g., \texttt{code}, \texttt{documentation}, \texttt{test}, \texttt{project}).
Each bar shows how many commits fall into each size bin.

The visual structure confirms the asymmetry between fixes and vulnerabilities.
Fixing commits overwhelmingly consist of small code patches.
Only a small fraction of fixes touch documentation or tests, and even those rarely exceed a few lines.
This pattern aligns with Linux kernel development practices, which emphasize narrowly-scoped, reviewable changes during backporting and maintenance.

In contrast, vulnerability-introducing commits are larger, more diverse, and span multiple file types.
Documentation changes are notably prevalent, with thousands of commits modifying 10 or more lines of documentation alongside significant code additions.
Contributions labeled as \texttt{project} or \texttt{other} (e.g., configuration files, build scripts) are also much more frequent in introducing commits.
This suggests that vulnerabilities often emerge as byproducts of ambitious, multi-faceted contributions such as new feature integrations or driver additions.

These findings suggest a key structural insight: \emph{vulnerability fixes are small and surgical, while the vulnerabilities themselves tend to be introduced in complex, wide-ranging commits that impact both code and auxiliary components}.
This observation has implications for vulnerability prevention: larger commits with broad file coverage, especially those touching documentation or project configuration, may warrant more intensive review and validation.

\begin{table*}[tbh]
  \centering
  \caption{Per-branch counts of CVSS\,v3.1 base metrics (per unique CVE).
\textmd{
    Metric labels follow CVSS\,v3.1 value codes: Availability (\texttt{A}), Attack Complexity (\texttt{AC}),
    Privileges Required (\texttt{PR}), Confidentiality (\texttt{C}), and Integrity (\texttt{I}): \texttt{H} = High, \texttt{L} = Low, \texttt{N} = None;
    User Interaction (\texttt{UI}): \texttt{N} = None, \texttt{R} = Required;
    Attack Vector (\texttt{AV}): \texttt{N} = Network, \texttt{A} = Adjacent, \texttt{L} = Local, \texttt{P} = Physical;
    Scope (\texttt{S}): \texttt{U} = Unchanged, \texttt{C} = Changed.
}}
  \label{tab:cvss-base-exploitability-by-lts}
  \begin{tabular}{lrrrrrrrr}
    \toprule
    & \textbf{AV} & \textbf{AC} & \textbf{PR} & \textbf{UI} & \textbf{S} & \textbf{C}& \textbf{I}& \textbf{A}\\
    \textbf{LTS} & \textbf{N/A/L/P} & \textbf{L/H} & \textbf{N/L/H} & \textbf{N/R} & \textbf{U/C} & \textbf{N/L/H} & \textbf{N/L/H} & \textbf{N/L/H} \\
    \midrule
    5.4   & 46 / 1 / \num{1438} / 3 & \num{1386} / 102 & 51 / \num{1421} / 16 & \num{1484} / 4 & \num{1487} / 1 & 939 / 18 / 531 & \num{1039} / 21 / 428 & 34 / 29 / \num{1425} \\
    5.10   & 64 / 1 / \num{2062} / 4 & \num{1983} / 148 & 75 / \num{2024} / 32 & \num{2127} / 4 & \num{2128} / 3 & \num{1400} / 25 / 706 & \num{1522} / 28 / 581 & 50 / 39 / \num{2042} \\
    5.15   & 56 / 2 / \num{2229} / 4 & \num{2115} / 176 & 67 / \num{2188} / 36 & \num{2288} / 3 & \num{2289} / 2 & \num{1533} / 29 / 729 & \num{1659} / 27 / 605 & 51 / 39 / \num{2201} \\
    6.1   & 59 / 3 / \num{1885} / 4 & \num{1816} / 135 & 67 / \num{1846} / 38 & \num{1948} / 3 & \num{1950} / 1 & \num{1299} / 20 / 632 & \num{1396} / 19 / 536 & 39 / 28 / \num{1884} \\
    6.6   & 65 / 5 / \num{2145} / 4 & \num{2076} / 143 & 79 / \num{2100} / 40 & \num{2214} / 5 & \num{2216} / 3 & \num{1511} / 21 / 687 & \num{1602} / 21 / 596 & 42 / 30 / \num{2147} \\
    6.12   & 1 / 0 / 603 / 1 & 582 / 23 & 4 / 600 / 1 & 605 / 0 & 605 / 0 & 389 / 3 / 213 & 423 / 0 / 182 & 0 / 1 / 604 \\
    \bottomrule
  \end{tabular}
\end{table*}

\subsection{Spot Check of Vulnerability-Inducing Commits}
\label{subsec:spot-check}

To better understand why many vulnerability-inducing commits in the Linux kernel are so large and complex, we performed a manual spot check of 20 large CVE-associated commits randomly drawn from several long-term support (LTS) branches.
During the analysis, three additional commits were added to capture edge cases not well represented in the initial sample, bringing the total to 23.
Each commit was independently reviewed by two researchers (an operating systems engineer and a security engineer) to ensure complementary perspectives on functionality and risk.
The detailed inspection notes are included in our replication package.

\paragraph{High-Level Observations.}
The vast majority of these commits fall into two categories (with one noticeable exception\footnote{Many CVEs in the kernel vulnerability {Git} repository are associated with the commit \href{https://git.kernel.org/pub/scm/linux/kernel/git/torvalds/linux.git/commit/?id=1da177e4c3f41524e886b7f1b8a0c1fc7321cac2}{1da177e4c3f41524e886b7f1b8a0c1fc7321cac2}.
That commit was authored by Linus Torvalds on April 16, 2005 during the migration from BitKeeper to Git with a note \say{\emph{I'm not bothering with the full history, even though we have it}.}}):
\begin{enumerate}
    \item \textbf{Full-feature additions}: new drivers, protocol implementations, or hardware enablement features. These patches are often thousands of lines long, spanning multiple files and subsystems.
    \item \textbf{Refactorings or rewrites}: renaming variables, reorganizing files, or updating infrastructure code. Although not functionally expansive, these changes still result in large diffs and can introduce subtle regressions.
\end{enumerate}

\paragraph{Root Causes of Commit Size.}
Large commits generally arise for two reasons.
First, maintainers often squash multiple smaller patches before integration to keep history concise, which reduces review granularity and complicates later forensic work.
Second, some contributions, especially new subsystems or drivers, are inherently monolithic and cannot be split without breaking interfaces or build dependencies.
As a result, a \say{vulnerability-inducing commit} is often only a coarse snapshot of a longer development process.
Additional context is often available in Linux Kernel Mailing List discussions, which record earlier revisions and review feedback.

\paragraph{Security Risk Origin.}
Most of these commits are not poorly written or unreviewed.
They are complex, well-engineered contributions that later revealed edge-case vulnerabilities.
Typical risk factors include user-kernel boundary handling (\texttt{sysfs}, \texttt{ioctl}, networking), concurrency and DMA management, and subtle behavioral drift introduced during refactoring.

\paragraph{Examples of CVEs that were introduced by large commits.}
\begin{itemize}
    \item \textbf{CVE-2025-22079}: Introduction of the entire OCFS2 file-system that consists of over 24,000 lines of code across 55 files.
    \item \textbf{CVE-2025-37970}: A fully-featured driver for the LSM6DSM MEMS sensor that includes I2C/SPI support.
    \item \textbf{CVE-2025-21941}: A refactoring of DRM display subsystem that introduced a vulnerability via variable renaming.
\end{itemize}

\paragraph{Implications.}
These findings suggest that many kernel vulnerabilities originate from the intrinsic complexity of large-scale system evolution rather than from negligence.
Vulnerabilities often emerge in broad, well-intentioned changes where reasoning and testing are difficult to exhaust.
Even refactorings and cleanup patches can introduce risk.
Maintainer squashing, while practical, reduces auditability and limits the precision of automated vulnerability analyses.

\subsection{CVSS Vectors Assigning Policy}%
\label{sub:CVSS vectors analysis}

\Cref{tab:cvss-base-exploitability-by-lts} shows the distribution of CVSS\,v3.1 base metric values across unique CVEs for each Linux LTS branch.
The distributions are strongly skewed toward specific values.
For example, most vulnerabilities require no user interaction (\texttt{UI=N}), affect systems over the network (\texttt{AV=N}), and assume low attack complexity (\texttt{AC=L}).
Similarly, the majority of CVEs have unchanged scope (\texttt{S=U}) and low or no required privileges (\texttt{PR=N/L}).

The Confidentiality (\texttt{C}), Integrity (\texttt{I}), impact metrics also exhibit skew, towards None (\texttt{C=N}, \texttt{I=N}).
Whole Availability (\texttt{A}) is predominantly High (\texttt{A=H}).
These imbalances are consistent across kernel versions and reflect common threat models in kernel-level vulnerabilities.

\begin{figure}[htpb]
    \centering
    \includegraphics[width=0.3\textwidth]{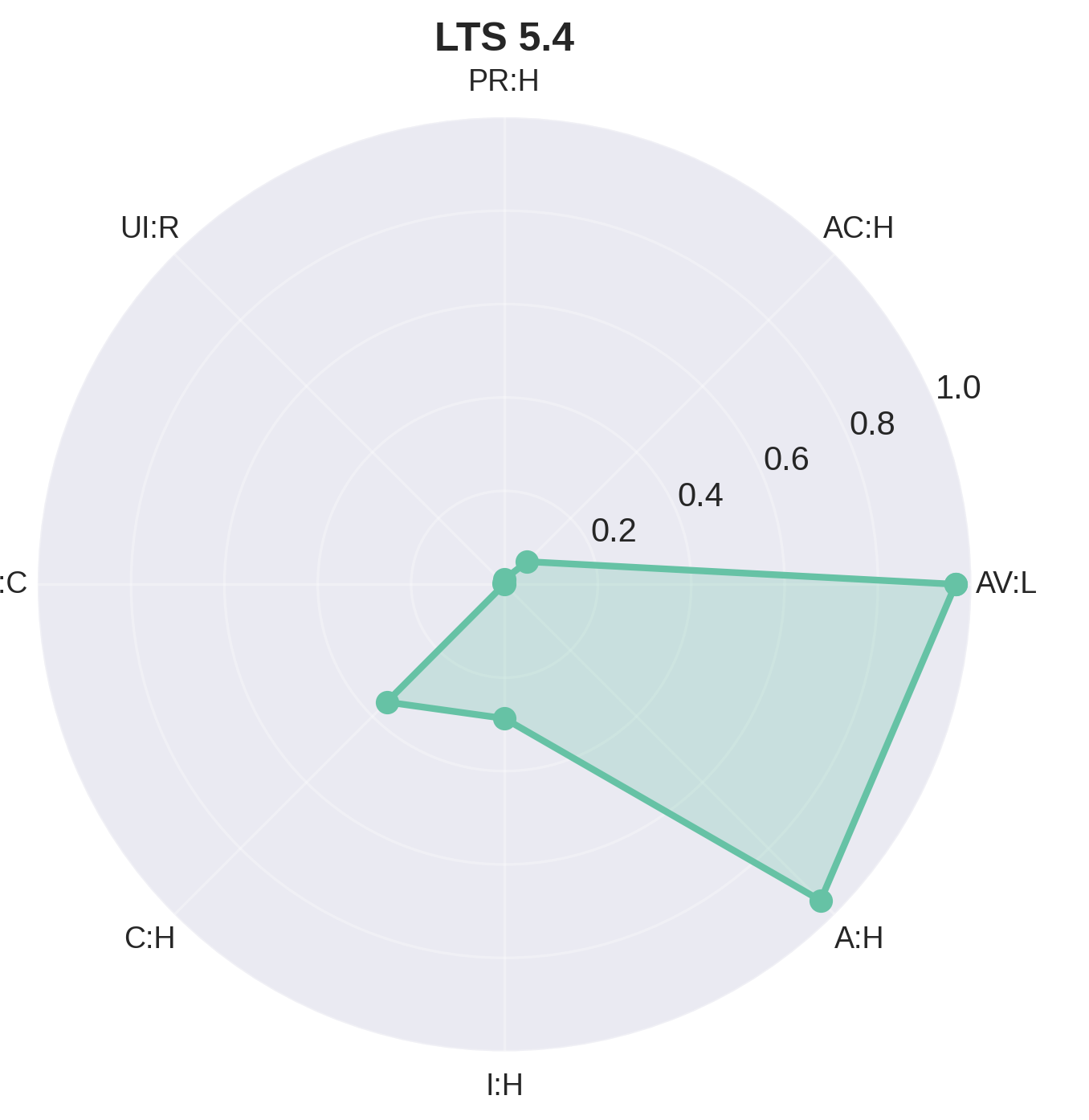}
    \caption{CVSS\,v3.1 distribution (radar plot) in Linux LTS 5.4 (similar across other LTS branches).}
    \label{fig:cvss_radar}
\end{figure}

Although the Linux kernel project manages its own vulnerability disclosures and fix commits,
the corresponding CVSS\,v3.1 vectors are typically assigned by external databases such as NVD or MITRE rather than by the kernel maintainers themselves.
Consequently, the reported CVSS values tend to follow standardized external scoring conventions instead of kernel-specific reassessment.
This is reflected in the radar profile of~\Cref{fig:cvss_radar},
where the prevalence of default combinations such as \texttt{AV=L}, \texttt{AC=L}, \texttt{PR=N},
and high-impact metrics (\texttt{C/I/A:H}) indicates consistent adoption of externally defined scoring patterns across kernel versions.

To quantify this uniformity, we applied association-rule mining (\texttt{apriori}, minimum support~0.95) to all unique CVSS vectors.
The dominant configuration-\texttt{AV=L}, \texttt{AC=L}, \texttt{PR=L}, \texttt{UI=N}, and \texttt{S=U}-appears in over 95\,\% of CVEs, with near-deterministic co-occurrence ($\texttt{AV=L}\Rightarrow\texttt{UI=N}$, confidence~$\approx0.999$).
This confirms that CVSS vectors in the kernel dataset follow a nearly fixed template, characterizing locally exploitable, non-interactive vulnerabilities with unchanged scope.

\subsection{Impact of CVSS Scores on Survival}
\label{sub:cve_score}

\begin{table}[htpb]
    \centering
\caption{CVE fix latency by CVSS\,v3.1 base metrics.
Each block shows a metric name in bold with Somers' $D_{xy}$ and its 95\,\% CI.
See~\Cref{tab:cvss-base-exploitability-by-lts} for label definitions.}
    \label{tab:survriskcvss}
\small
\begin{tabular}{lrrrrrr}
\toprule
Label & Count & Min & Median & Mean & Max & Std \\
\midrule
\multicolumn{7}{c}{ \textbf{Availability} \Dxy -0.0, \CI~[-0.0, -0.0] } \\
\midrule
H & \num{19036} & \num{3} & \num{1929} & \num{2360} & \num{7396} & \num{1804} \\
L & \num{258} & \num{22} & \num{2189} & \num{2454} & \num{5921} & \num{1844} \\
N & \num{399} & \num{23} & \num{2052} & \num{2190} & \num{5672} & \num{1567} \\
\midrule
\multicolumn{7}{c}{ \textbf{Attack Complexity} \Dxy -0.01, \CI~[-0.01, -0.01] } \\
\midrule
H & \num{1435} & \num{9} & \num{2164} & \num{2507} & \num{7220} & \num{1815} \\
L & \num{18258} & \num{3} & \num{1920} & \num{2346} & \num{7396} & \num{1799} \\
\midrule
\multicolumn{7}{c}{ \textbf{Attack Vector} \Dxy 0.0, \CI~[0.0, 0.0] } \\
\midrule
A & \num{8} & \num{83} & \num{2064} & \num{2249} & \num{4784} & \num{1788} \\
\midrule
\multicolumn{7}{c}{ \textbf{Confidentiality} \Dxy -0.04, \CI~[-0.04, -0.04] } \\
\midrule
H & \num{6372} & \num{3} & \num{2120} & \num{2524} & \num{7271} & \num{1827} \\
L & \num{202} & \num{23} & \num{2043} & \num{2216} & \num{6035} & \num{1484} \\
N & \num{13119} & \num{3} & \num{1861} & \num{2279} & \num{7396} & \num{1786} \\
\midrule
\multicolumn{7}{c}{ \textbf{Privileges Required} \Dxy 0.0, \CI~[0.0, 0.0] } \\
\midrule
H & \num{253} & \num{143} & \num{1627} & \num{2050} & \num{5793} & \num{1382} \\
L & \num{18822} & \num{3} & \num{1941} & \num{2364} & \num{7396} & \num{1808} \\
N & \num{618} & \num{22} & \num{1916} & \num{2292} & \num{5672} & \num{1697} \\
\midrule
\multicolumn{7}{c}{ \textbf{User Interaction} \Dxy 0.0, \CI~[0.0, 0.0] } \\
\midrule
N & \num{19654} & \num{3} & \num{1931} & \num{2358} & \num{7396} & \num{1801} \\
R & \num{39} & \num{321} & \num{2464} & \num{2226} & \num{4784} & \num{1227} \\
\midrule
\multicolumn{7}{c}{ \textbf{Scope} \Dxy -0.0, \CI~[-0.0, -0.0] } \\
\midrule
C & \num{16} & \num{2200} & \num{2200} & \num{2529} & \num{4831} & \num{899} \\
U & \num{19677} & \num{3} & \num{1931} & \num{2357} & \num{7396} & \num{1801} \\
\midrule
\multicolumn{7}{c}{ \textbf{Severity} \Dxy -0.01, \CI~[-0.01, -0.01] } \\
\midrule
U & \num{13164} & \num{4} & \num{2037} & \num{2541} & \num{7396} & \num{1978} \\
L & \num{119} & \num{199} & \num{1562} & \num{2089} & \num{5921} & \num{1739} \\
M & \num{13417} & \num{3} & \num{1881} & \num{2291} & \num{7396} & \num{1782} \\
H & \num{6049} & \num{3} & \num{2066} & \num{2518} & \num{7271} & \num{1840} \\
C & \num{108} & \num{97} & \num{2115} & \num{1934} & \num{3640} & \num{1145} \\
\bottomrule
\end{tabular}
\end{table}

The results in~\Cref{tab:survriskcvss} show that CVSS~v3.1 base metrics, both individual vector components and derived severity levels, have a negligible association with the time required to fix Linux kernel~CVEs.
\emph{Note:} Only observations with specific CVSS\,v3.1 metric are included, and the population size is therefore constrained by this subset.

Across all metrics, Somers'~$D_{xy}$ values are effectively zero, with very narrow bootstrapped 95\,\% confidence intervals collapsing to~$[0,0]$, indicating no detectable monotonic relationship between CVSS categories and fix latency.
\emph{Note:} Although several CVSS components are strongly imbalanced, the consistently near-zero Somers'~$D_{xy}$ values and tight confidence intervals suggest that the absence of association is robust rather than a sampling artifact. Still, such imbalance limits the sensitivity of rank-based measures and should be considered when interpreting effect size.

As a robustness check, we additionally applied global log-rank tests to compare time-to-fix distributions across CVSS categories.
While several CVSS components yield statistically significant results, these effects are driven by large sample size and pronounced group imbalance.
Consistent with this observation, effect sizes measured via Somers'~$D_{xy}$ remain close to zero, indicating no meaningful or monotonic relationship between CVSS metrics and fix latency.
Thus, log-rank significance reflects statistical detectability rather than practical relevance.

This skew-driven by dominant categories like \texttt{Low} attack complexity, \texttt{None} user interaction, and \texttt{High} availability impact, reduces the discriminative capacity of CVSS metadata for vulnerability triage.
Even when aggregating to severity levels, no signal emerges: \say{Critical} CVEs have a higher median time-to-fix (\num{2115} days) than \say{Low} ones (\num{1562} days), despite their nominally higher urgency.
These findings suggest that technical and organizational factors, such as subsystem ownership, patch complexity, backport feasibility, and maintainer workload, play a far greater role in patching latency than CVSS-based assessments.

\subsection{Age of LTS Kernels}
\label{sub:Fixes across LTS versions}

For RQ3, we compare the full time-to-fix distributions across LTS branches rather than only summary statistics.
This is important because fix latency is heavily right-skewed and includes censoring (some CVEs are not observed as fixed within the study window), making survival analysis a natural fit.
We therefore report Kaplan--Meier curves, log-rank tests, and a concordance-based effect size for the version-age gradient.

\begin{figure}[htpb]
    \centering
    \includegraphics[width=0.5\textwidth,scale=0.7]{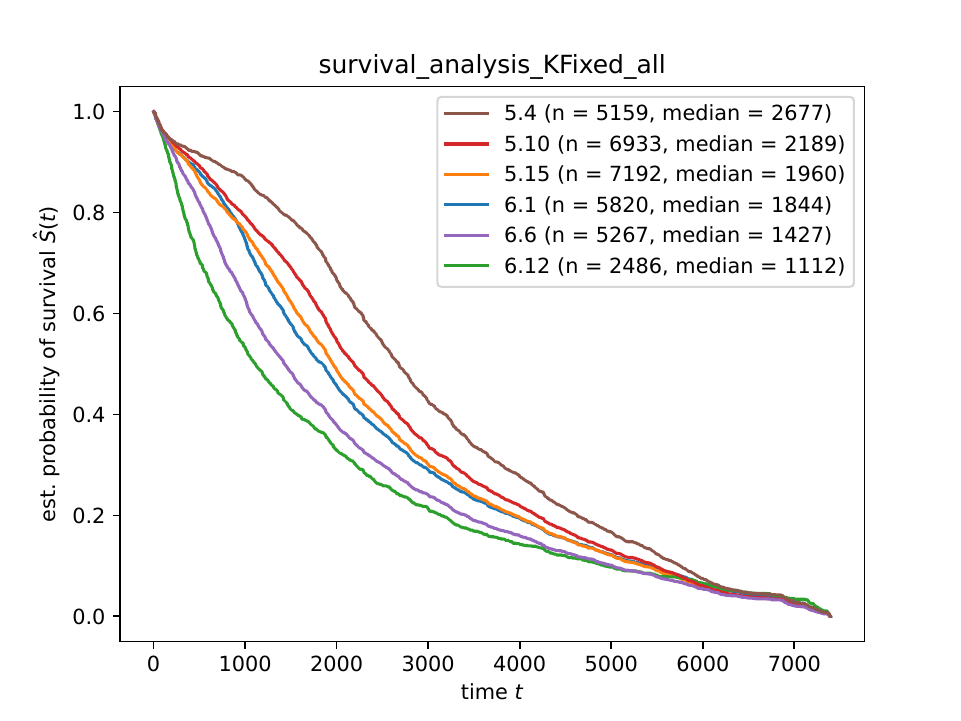}
    \caption{Survival analysis of CVE fix latency across Linux LTS versions.
    Newer versions (e.g., 6.12, 6.6) exhibit shorter median fix times compared
    to older lines (e.g., 5.10, 5.4).
}
    \label{fig:img-survival_analysis_KFixed_all-pdf}
\end{figure}

Across the \num{32857} CVE-LTS observations in our dataset, the overall survival median fix time is \num{1976} days ($\approx$5.4 years).
\Cref{fig:img-survival_analysis_KFixed_all-pdf} presents the Kaplan-Meier survival curves, which indicating that \textbf{newer kernel versions exhibit a higher hazard of receiving a fix compared to older ones}.
For example, the median fix time in branch~6.12 is about \num{1112} days ($\sim$3.0 years), whereas in the oldest analyzed branch~5.4 it extends to nearly \num{2677} days ($\sim$7.3 years).

\begin{table}[tbh]
  \centering
  \caption{Survival analysis summary across all LTS branches. LTS version acts as a risk factor: newer versions tend to receive fixes faster.}
  \label{tab:lts-fix-time-survival}
  \begin{tabular}{lrrrrrr}
    \toprule
    LTS & Count & Median & Min & Max & Std & Mean \\
    \midrule
    5.4 & \num{5159} & \num{2677} & \num{5} & \num{7396} & \num{1839.2} & \num{2981.40} \\
    5.10 & \num{6933} & \num{2189} & \num{3} & \num{7396} & \num{1830.29} & \num{2593.47} \\
    5.15 & \num{7192} & \num{1960} & \num{3} & \num{7396} & \num{1835.91} & \num{2414.26} \\
    6.1 & \num{5820} & \num{1844} & \num{3} & \num{7396} & \num{1870.43} & \num{2364.54} \\
    6.6 & \num{5267} & \num{1427} & \num{3} & \num{7396} & \num{1836.36} & \num{2047.33} \\
    6.12 & \num{2486} & \num{1112} & \num{6} & \num{7396} & \num{1919.68} & \num{1851.21} \\
    \bottomrule
  \end{tabular}
\end{table}

\Cref{tab:lts-fix-time-survival} provides per-branch survival estimates.
Median latency grows steadily with branch age: from \num{1112} days in 6.12 and \num{1427} days in 6.6, through \num{1844} days in 6.1 and \num{1960} days in 5.15, up to \num{2189} days in 5.10 and \num{2677} days in 5.4.
Although the distributions overlap (substantial standard deviations of $\sim$\num{1800} days across all branches), the ordering of medians reveals a consistent gradient.

When branches are sorted from newest to oldest, the concordance index is $D_{xy}=0.13$ with a narrow 95\,\% confidence interval, confirming that LTS version is a modest risk factor in survival analysis terms.
In other words, kernel version age provides above-chance discriminatory signal for time-to-fix, although the effect size remains moderate.

\section{Discussion}
\label{sec:discussion}

\subsection{Commit Anatomy}
\label{sub:discussion-commit-anatomy}

Our analysis of over \num{30000} annotated commits reveals a pronounced structural asymmetry between vulnerability-introducing and fixing commits.
Fixes are typically narrow in scope: more than \num{11000} touch fewer than five lines of code and almost all modify a single file.
Changes outside the core codebase (documentation, configuration, or tests) are rare and usually peripheral.
This pattern aligns with long-standing kernel development norms that favor small, reviewable patches, particularly within stable and long-term support branches.

In contrast, vulnerability-introducing commits are substantially larger and more heterogeneous.
They often modify dozens of files and thousands of lines, extending beyond code to include documentation, configuration, and build scripts.
Documentation updates in particular tend to scale with the size of the contribution, reflecting integration of new drivers, subsystems, or architectural features.

A qualitative spot check of 23 large CVE-associated commits supports these findings.
Manual inspection showed that many vulnerability-inducing commits are not carelessly written but rather represent complex, well-reviewed engineering efforts.
Those commits are often multi-patch features that were later squashed into single commits for integration.
Such practices simplify project history but obscure incremental review context, making forensic reconstruction harder.
In several cases, vulnerabilities arose from the introduction of new user-kernel interfaces (e.g., \texttt{sysfs}, \texttt{ioctl}, or networking) or from changes introduced during large-scale refactoring.
These observations reinforce that vulnerability-inducing commits often capture only a coarse snapshot of a longer development process, where the actual defect may have been seeded over multiple revisions.

Together, these results suggest that vulnerabilities seldom originate from isolated mistakes.
They are more likely side effects of large-scale, legitimate engineering efforts, while fixes, once the root cause is known, tend to be concise, localized, and easy to apply.

\begin{tcolorbox}[colback=gray!10, colframe=gray!50,
title=RQ1: What does the anatomy of vulnerability-introducing and fixing commits reveal about the cost of addressing vulnerabilities?]
Large, multi-file commits frequently introduce vulnerabilities as by-products of complex feature growth or architectural change, while fixes remain small and surgical.
This asymmetry implies that the effort of repair lies not in implementing the patch but in identifying, diagnosing, and coordinating its integration across kernel branches.
\end{tcolorbox}

\subsection{CVE Metadata and Patch Prioritization in the Linux Kernel}
\label{sub:discussion-cvss-metadata}

Our analysis suggests that CVSS metadata, both the overall severity score and the individual base vector components, has minimal influence on how quickly Linux kernel vulnerabilities are fixed.
Although CVSS aims to capture exploitability and impact, these factors show no systematic relationship with vulnerability lifetime across Linux LTS branches.

Several factors likely contribute to this outcome.
First, the distribution of CVSS vector values is heavily skewed: most kernel CVEs are marked as low-complexity, requiring no user interaction, and having a high impact on at least one dimension.
This distribution limits the resolution of CVSS in differentiating risk across the kernel vulnerabilities.

Second, Linux kernel patching practices are shaped more by social and technical dynamics, such as developer responsibility, subsystem structure, and backport difficulty, rather than by external severity scoring.
As a self-organized open-source project with long-lived branches and decentralized governance, the kernel community operates outside rigid triage pipelines.

Third, although Linux now acts as its own CNA, the CVSS vectors assigned by NVD remain formulaic.
Although internally consistent, they vary too little across CVEs to reflect practical differences in urgency or technical risk.
As a result, CVSS provides limited guidance for either prioritization or retrospective analysis.

\begin{tcolorbox}[colback=gray!10, colframe=gray!50, title=RQ2: What role do CVSS scores play in the speed of vulnerability fixing in the Linux kernel?]
CVSS metadata, including both base vectors and severity scores, has a negligible influence on vulnerability fix latency. Patch timing appears largely independent of standardized CVSS risk assessments, and is instead shaped by project-internal factors such as maintainership, backport complexity, and subsystem responsibility.

\textcolor{red!70!black}{\textbf{Additional observation:}}
Beyond its lack of predictive power, CVSS scoring in the Linux kernel offers limited descriptive value: most vectors follow fixed templates with little variation, reflecting the kernel's narrow vulnerability profile and standardized CVE evaluation policy.
\end{tcolorbox}

\subsection{Kernel Version and Fix Latency}
While our analysis confirms a systematic risk gradient across kernel versions, it is important to clarify its statistical meaning.
The concordance index ($D_{xy}=0.13$) with a narrow 95\,\% confidence interval indicates that the kernel version provides discriminatory power above random chance, but the effect size remains modest.
This finding suggests that while version age is predictive, it is not a strong discriminator on its own.

Nevertheless, the practical implication is clear: median fix times consistently
increase with branch age (from $\sim$\num{1100} days in 6.12 to $\sim$\num{2700} days
in 5.4).
This systematic gradient aligns with the community knowledge that
backporting to older branches tends to be slower and more complex, and it highlights a
security trade-off.
Organizations that remain on older LTS kernels---often due
to certification, compliance, or stability requirements---face longer exposure
windows for known vulnerabilities.
Even if kernel age is not the sole driver of
fix latency, its consistent predictive signal makes it an important risk factor
to consider in long-term maintenance and security policies.

\begin{tcolorbox}[colback=gray!10,colframe=gray!50,title=RQ3: How does the kernel version age influence the speed of vulnerability fixes?]
Kernel version age modestly but consistently influences the speed of
vulnerability fixes. Median fix times grow steadily from newer to older LTS
branches (about \num{1100} days in 6.12 versus nearly \num{2700} days in 5.4).
The concordance index ($D_{xy}=0.13$) indicates limited predictive power, but
with a stable confidence interval above chance, kernel age can be regarded
as a meaningful, though not dominant, risk factor for delayed patching.
\end{tcolorbox}

\section{Threats to Validity}
\label{sec:threats}

Our analysis of CVE lifetimes and patch characteristics in the Linux kernel is subject to several limitations~\cite{shull_guide_2008}.
We discuss them below in terms of internal, construct, and external validity.

\paragraph*{Internal validity.}
While we make extensive use of commit metadata (timestamps, tags, and CVE annotations), inaccuracies in commit labeling or backporting practices could affect our measurements.
Although the Linux kernel CNA typically provides accurate vulnerability-introducing and fixing commits, CVEs can be assigned retroactively, and some CVEs may have missing or inconsistent metadata.
We mitigate this threat by filtering for CVEs with both start and end timestamps.
Additionally, one CVE may be fixed multiple times across LTS branches or even within the same branch, possibly due to incorrect, partial, or staged fixes.
We treat each such fix as a valid independent observation, but acknowledge potential redundancy.

An outlier case we did not investigate is other entities  (e.g., distribution vendors, security researchers, third-party CNAs) assigning CVE IDs for vulnerabilities found in the official kernel codebase.
The accepted industry practice is for anyone else to coordinate with the Linux kernel CNA because it is an authoritative source. However, nothing formally prevents third parties from submitting the requests directly to NVD and NIST.
We observe the Linux kernel CNA is inspecting all those CVEs and rejecting or withdrawing them if needed (e.g., CVE-2025-0927\footnote{\url{https://nvd.nist.gov/vuln/detail/CVE-2025-0927}}).

\paragraph*{Construct validity.}
We define CVE lifetime as the time between vulnerability introduction and fix in a given kernel branch.
This does not always match when a vulnerability was publicly disclosed or assigned a CVE.

We use introduction time rather than CVE reporting time as our main temporal reference.
This is a construct validity trade-off.
In the Linux kernel, CVE reporting is often delayed, retroactive, or driven by maintenance and backporting needs rather than initial exposure.
In recent years, CVEs have frequently been assigned to defects introduced long before disclosure, mainly to motivate patch propagation.
As a result, reporting time does not reliably reflect when a kernel version first became vulnerable.

Using introduction time better approximates actual exposure within a branch, but reduces comparability with studies based on disclosure timelines.
Our conclusions focus on relative lifetimes and patching behavior, which are robust to this choice.

Finally, our patch size and commit anatomy analysis is based on static line-level changes and metadata, and may not fully capture semantic complexity or development effort.

\paragraph*{External validity.}
Our results pertain to the LTS branches of the Linux kernel and may not generalize to mainline development or other open-source operating systems and projects.
The kernel community has a unique development culture, including decentralized maintainership and long-term backporting practices.
These factors influence how and when fixes are applied, possibly in ways not captured by CVE metadata.
Thus, care must be taken when extrapolating our results to other ecosystems.

\section{Conclusions and Future Work}
\label{sec:conclusion}
We presented an empirical study of Linux kernel CVEs since the project
became a CNA. Our analysis highlights the limited role of CVSS metadata
in predicting fix latency, structural insights from commit anatomy,
and the effect of kernel version age on patch speed.
Future work includes extending commit annotation,
cross-project comparisons, and predictive modeling of vulnerability resolution.

\subsection{Implications to the Industry}

\begin{flushright}
    \emph{\say{Nobody who relies on backporting fixes to a non-mainline kernel will be able to keep up with this CVE stream. Any company that is using CVE numbers to select kernel patches is going to have to rethink its processes.}}
\end{flushright}
\begin{flushright}
--- Jonathan Corbet~\cite{lwn-cve-article}.
\end{flushright}

\subsubsection{Move to the latest kernel versions}

The increased inflow of CVEs has drastic implications for the industry.
Since becoming the CNA, the Linux kernel project has published on average $14$ new kernel CVEs per day~\cite{cves-per-day-script}.
In comparison, between 2006 and 2018, approximately $84$ CVEs per year impacted the Linux kernel~\cite{tux_care}.
A simple interpretation of our findings from~\Cref{sub:Fixes across LTS versions} is that CVEs are fixed more quickly in the newer kernel versions.
The backporting of patches to older LTS kernels is done by either the Linux kernel security team or the maintainers of the impacted subsystems.
Given the high consequences of defects in the kernel code, each backport requires a detailed understanding of the code changes and focused testing.
The increased number of these activities takes a toll on a limited number of maintainers.
The burnout of open-source software maintainers is a known problem~\cite{maintainer-burnout}.
In recent years, the topic has also become relevant to the Linux kernel community~\cite{lwn-maintainer-burnout,zdnet-maintainer-burnout}.

The direction to use the more recent kernel versions is not new.
For years, the guidance from the Linux community has been to upgrade to the \emph{latest} stable or LTS release~\cite{use-the-latest-kernel}.
Those recommendations align with the major open-source projects and companies embracing the \say{upstream-first} approach to kernel development~\cite{meta-upsteam-first,android-upstream-first}.

\subsubsection{Limited application of NVD ratings}

The findings in~\Cref{sub:discussion-cvss-metadata} indicate that the Linux kernel community does not backport the fixes according to the NVD severity ratings.
This finding may appear counterintuitive to an observer outside the kernel community.
However, this behavior aligns with the long-held belief in the kernel community.
Due to the diverse ways in which the Linux kernel is utilized, it is not possible to provide a reliable unified CVE rating~\cite{gkh-talk-cves-are-alive}.
Each organization that provides or uses a custom kernel will have to perform its own triage process that is tailored to its specific execution environment and security requirements.

\subsubsection{European Union Cyber Resilience Act}

Another challenge the industry faces is complying with the {E}uropean {U}nion {C}yber {R}esilience {A}ct (CRA)~\cite{eu-cra}.
The CRA establishes mandatory security requirements on the software intended for the European market.
According to the CRA, software manufacturers are legally responsible for the security of their products.
A part of that responsibility includes reporting and patching the known vulnerabilities.
The reporting timelines are relatively aggressive.
For example, CRA requires software manufacturers to report actively exploited security defects to the European Union Agency for Cybersecurity within 24 hours of their discovery~\cite{eu-cra-article-14,cra-reporting-24-hours}.

\subsection{Future Directions in Vulnerability Management}

Our informal discussions with industry practitioners about the changes in the Linux kernel CVE ecosystem confirm the following recommendations from the kernel community.

\begin{quote}
    \say{\emph{In order to run the safest kernel, one must choose to either update to latest stables and do all the work needed to get it running in production, or one must triage every fix to find only those that need to be applied based on one's threat model.}
\end{quote}

\begin{quote}
    \emph{Either way is likely a lot of work, but one must figure out which is the least amount of work, and then do it.}
\end{quote}

\begin{quote}
    \emph{There isn't another path to running a kernel with the flaws fixed.}}
    \begin{flushright}
        \emph{-- Kees Cook}~\cite{gkh-talk-2024-suddenly}
    \end{flushright}
\end{quote}

We do not expect the majority of Linux kernel consumers to migrate to the latest stable kernels rapidly.
One of the kernel community's foundational principles is that \say{WE DO NOT BREAK USERSPACE!}~\cite{linus-do-not-break-userspace}.
However, while inspirational, this goal is challenging to achieve given the numerous possible scenarios in kernel usage.
That implies the continued demand for rapid backporting of patches to the older kernel versions.

Given the limited number of maintainers available, the increasing number of kernel CVEs, and the overall focus on AI- and LLM-assisted software development, we anticipate a greater emphasis on automated backporting.
Researchers have investigated automated backporting in the past~\cite{shariffdeen-kernel-backport,yang-oss-backport,serrano2020spinfer}.
However, this research predates when the Linux kernel project became CNA.
We observe active development of tools, such as \say{\textsc{AUTOSEL}: Modern AI-powered Linux Kernel Stable Backport Classifier} to handle the increase in patch backporting~\cite{autosel-kernel-backport}.
We do not think that the current approach of 3--4 kernel developers reviewing each stable kernel commit is sustainable in the long term~\cite{gkh-talk-2024-suddenly}.

\subsection{Final Remarks}

Our results provide a data-driven confirmation of what many Linux kernel developers have long suspected but rarely quantified.
Fixes for vulnerabilities arrive faster in newer kernel branches.
This finding implies that moving to recent LTS or stable kernel versions significantly reduces the exposure to vulnerability windows.
For the industry users, this finding translates into a clear operational message: \emph{upgrading to newer kernels is not merely a matter of feature access, but of measurable security advantage and reduction in software maintenance costs}.

In the context of other projects, the CVSS metadata plays a central role in vulnerability management.
We find that it has limited predictive power in the Linux kernel ecosystem.
In retrospect, this observation may be by design.
The NVD applies the CVSS templates consistently, emphasizing uniformity and transparency over fine-grained prioritization.
We recommend that the CVSS values should not be interpreted as signals of patch urgency, but rather as formal records of disclosure.
Organizations relying on Linux kernel CVSS metadata for triage should reassess their processes accordingly, taking into account their specific kernel configurations, threat models, and deployment contexts.

Finally, the sustainability of open-source security cannot be separated from the people who maintain it.
The increasing volume of security reports and the growing compliance expectations from industry are becoming unsustainable for unpaid volunteers~\cite{TriagingSecurityIssues2025,zdnet-maintainer-burnout}.
If companies depend on timely fixes or prioritization of specific vulnerabilities, they should contribute to open-source projects either through direct engineering support or financial compensation.
Without such reciprocity, the current model of volunteer-driven vulnerability management risks becoming untenable.
Security in open source is ultimately a shared responsibility between maintainers and the industries that depend on their work.

In short, our study empirically grounds several long-held community intuitions: staying current matters, and CVSS does not capture the actual patching dynamics of the Linux kernel.

\begin{acks}
The authors thank Matthew S.~Wilson for their feedback on the early draft, and Thor Skaug for supporting this line of research.
\end{acks}

\newpage
\balance

\bibliographystyle{ACM-Reference-Format}
\bibliography{paper}

\end{document}